## Transient Liquid Phase Bonding of NiTi Using Cu- and Nb-base Interlayers

Zhaoxi Cao[1], Samuel Price[1], Alessandra Crippa[2], John P. Reidy[1], Gianna M. Valentino[2], Ian McCue[1*]

[1]*Department of Materials Science and Engineering, Northwestern University, Evanston, IL 60208, USA*

[2]*Department of Materials Science and Engineering, University of Maryland, College Park, MD, 20742, USA*

*Corresponding author: ian.mccue@northwestern.edu*

## Abstract

Transient liquid phase (TLP) bonding was examined as an approach for joining NiTi to achieve a high joint efficiency while minimizing chemical variance within the joint region. Two bonding interlayer chemistries (Cu-base and Nb-base) were identified by screening thermodynamic criteria for TLP in ternary alloys using the CALPHAD method. These two systems were then experimentally evaluated with respect to their impact on solidification kinetics, microstructure in the joint region, and performance during quasistatic and cyclic tensile loading. For both interlayer chemistries, the composition profile and microstructure in the joint region confirmed an isothermal solidification mechanism. In addition, the joints were found to be fully dense and contain at most 1.2% intermetallic phases. Tensile testing showed excellent load transfer across the joints with approximately 4% recoverable strain and martensite onset stresses reaching 94% and 89% of the unbonded, annealed NiTi values for Cu-base and Nb-base interlayers, respectively. Lastly, a stable superelastic response was observed under cyclic loading for both bond chemistries, with spatial variation in the strain evolution linked to enhanced stiffness and hardness in the joint region arising from the substitutional Cu and Nb solutes, as confirmed via nanoindentation. This study demonstrates that TLP bonding of NiTi can produce high-strength and nearly intermetallic-free joints without sacrificing functional performance, such as the superelastic response.

# 1. Introduction

Shape memory alloys (SMAs) are of interest for aerospace applications due to their excellent mechanical behavior, corrosion resistance, and functional properties [1, 2]. In particular, NiTi (a roughly equiatomic alloy) is the most widely used SMA because of its superior shape memory and superelastic effects – capable of restoring strains up to 8%. These physical and functional properties are well-suited for self-deploying actuators and other remote environment applications, which often require NiTi to be joined to other materials [3-5]. However, the unique superelastic and shape memory properties of these materials are highly sensitive to compositional variance (at the 100-ppm level), impurities (e.g., C and O), and the underlying microstructure, all of which can be altered during welding and other joining processes [1, 2]. Additionally, NiTi's ordered crystal structure is limited to a narrow composition range, and deviations of a few atomic percent lead to formation of intermetallic phases.

Due to these complications, NiTi components in service tend to be mechanically joined using bolts and/or fasteners [5, 6]. For instance, the superelastic tire developed by researchers at NASA Glenn relies on NiTi wires attached to a metallic tire rim via mechanical crimping [7-9]. Although this approach is straightforward for assembly, the joint can be heavy and bulky. Furthermore, mechanical joints do not provide smooth load transfer, which leads to limited shape memory/superelastic actuation stroke, low joint pull-out force, and poor cycle life [6]. This need for robust NiTi joining solutions that do not degrade its performance has spurred a wealth of research on the topic [10-15].

Only a few techniques have been able to produce high joint efficiencies (>80%), with laser welding receiving the most attention due to its ease of use and widespread industry adoption [6, 16-22]. However, fusion-based processes are challenging to optimize for shape memory/superelastic performance. Selective elemental evaporation is a major concern – even for the short melting and solidification timescales during laser scanning (20 mm/sec) – with one study reporting substantial variation in the transformation temperature across the heat-affected zone [23]. Solid-state techniques like friction stir welding (FSW) and explosive joining can circumvent the issues caused by melting and solidification, but can have poor cyclic performance (FSW specifically) and require operator line of sight (explosive joining) [24-26].

Diffusion bonding is typically employed when a material is sensitive to changes in chemistry, susceptible to cracking, or the service environment is too harsh for adhesives [27-29]. However, successful diffusion bonding needs labor-intensive surface preparation, applied pressure, high temperature, and a long bonding time. These drawbacks are largely overcome by a subset of diffusion bonding known as transient liquid phase (TLP) bonding. TLP bonding relies on an interlayer (i.e., foils or powder containing pastes) with solute elements that promote the formation of a thin liquid between the two materials to be joined, and solidification is controlled by solid-state diffusion of the solutes through the substrate(s). TLP bonding is amenable to a range of joint geometries while requiring minimal operator intervention and can be used to attach complex components through a flexible interlayer. Commercially, TLP processes are used for the fabrication and repair of Ni-base superalloy turbine blades, with joining times of ~3 hours and little-to-no applied pressure [30, 31].

TLP joining of NiTi has been previously demonstrated by Grummon et al., where the reaction of Nb and NiTi formed a NiTiNb eutectic liquid [32, 33]. Next to explosive joining, this work produced one of the highest reported joint efficiencies for NiTi. However, this study is more accurately described as reactive brazing. Rather than holding the joint at an elevated temperature for an extended time – allowing Nb to diffuse into the adjacent NiTi substrates and drive solidification – the assembly was quenched and the liquid solidified into a eutectic microstructure consisting of Nb-rich and NiTi-rich phases. This NiTiNb liquid has been used for efficient sintering of powder NiTi compacts [34, 35], but there has been no definitive study on TLP processes in NiTi and several outstanding questions remain regarding identification of other viable TLP solutes, their solidification behavior, and their subsequent impact on the mechanical and functional properties of NiTi in the joint region.

In this study, we assessed the feasibility of TLP bonding for similar joining of superelastic NiTi. First, solute additions to NiTi were screened via CALPHAD to identify ternary liquids that are in equilibrium with NiTi (i.e., below the melting point of NiTi). From this screening, eight promising ternary systems were identified, and descriptors for TLP figures of merit were used to select the Ni-Ti-X (X = Cu or Nb) systems that minimize the solidification time. Next, these systems were evaluated experimentally. The solidification timescales and rate-limiting behavior of Cu-base and Nb-base liquids in contact with NiTi were quantified, and an isothermal solidification mechanism was confirmed. Macroscopic joints were then fabricated from these two compositions to understand the role of solutes on the joint microstructure formed within the joint region. Finally, joined specimens were mechanically evaluated via quasistatic and cyclic tensile loading. Both solutes were found to produce high joint efficiencies, and the joints survived 10 cycles without failure. Nanoindentation identified a local increase in elastic modulus and hardness in the joint region arising from the dilute (<5 at.%) Cu and Nb solutes. This work demonstrates the effectiveness of TLP bonding to produce dense and strong joints in NiTi without sacrificing the performance of the joined material.

# 2. Methods

## 2.1. CALPHAD-guided Selection of Bonding Interlayers

The CALPHAD calculations used in this framework were performed by Thermo-Calc software version 2023b using the TC-Python module and TCHEA5 and MOBHEA3 database [36, 37]. As an initial screening, Ni-Ti-X ternaries (where X is every element in the HEA database) were evaluated using single-equilibrium calculations from 800-1400K in 100K increments. Ternary systems of interest were identified based on whether a liquid phase exists in equilibrium with the B2 NiTi phase. Isothermal phase diagram calculations were performed to determine the two-phase region with higher resolution and the concentration range for the third element along the liquidus and solidus.

While binary TLP permits analytical solutions for the time required for isothermal solidification [38], the ternary case does not as it is complicated by multicomponent diffusion and non-fixed interfacial compositions [39]. Therefore, exact prediction of the isothermal solidification time in these ternary systems would require numerical modeling. Instead, we implement a heuristic approach under the assumption that high solute solubility and diffusivity for the ternary element should promote faster solidification.

## 2.2. Specimen Acquisition and Preparation

Two specimen geometries were used in this study, plates for solidification and rods for joining, which are labeled and shown in Figure 1. Both geometries were prepared via wire EDM from NiTi alloy rod feedstock purchased from Kellogg Research Lab. The vendor reported a nominal transition temperature of -15 °C associated with this composition. LECO was used as a third-party service to confirm the composition and quantify impurities using ICP-OES, combustion furnace, and inert gas fusion methods. The measured composition was (in wt.%): 56.1 Ni; 43.77 Ti; 0.027 O; and 0.063 C.

For the solidification study, Figure 1(a), 10x10x13 $mm^3$ NiTi prisms with cylindrical holes (5 mm in diameter and height) were filled with powder from one of the two TLP bonding systems (Cu- or Nb-base). For the Cu-base system, Ti and Cu powders were mixed at a ratio corresponding to a target composition of Ti-60 at.% Cu. For the Nb-base system, pure Nb powder was used. The weight of the powders filled into the cylindrical holes was between 0.15-0.19 g. After filling, the specimens were sealed inside quartz ampoules along with Ti shim as an oxygen getter. The quartz ampoules were evacuated to ~$10^{-1}$ torr and backfilled with argon with the pressure calculated to reach 0.8 atm at the heat-treatment temperature. The NiTi-Cu setup was heat-treated at 970 °C, 1120 °C, and 1170 °C for 10 min to 24 h. The NiTi-Nb setup was heat treated at 1170 °C, 1200 °C, and 1240 °C for 30 min to 25 h. The samples were all water quenched after heat treatment.

TLP bonded joints were prepared using NiTi rods with a diameter of 25 mm and 38 mm in length, with bond surfaces mechanically ground (SiC papers to 1200 grit) to remove surface oxides and reduce surface roughness. Figure 1(b) illustrates a holder design that ensures the two NiTi rods to be joined remain in place during TLP bonding. A 10 mm-thick TZM (Ti-Zr-Mo) plate was used to align the two rods during the bonding process, and alumina crucibles were used to cover the rods to minimize contamination during bonding. Figure 1(c) illustrates the interlayer setup for both bonding chemistries. Foils were used instead of powder-based pastes after preliminary studies showed that foils produced joints with lower porosity. Ti,

Cu, and Nb metal foils (99.9 wt.% purity or higher) were used as bonding interlayer materials with thicknesses of 11.4 µm, 10.0 µm, and 50.0 µm, respectively. All the foils were purchased from Thermo Fisher Scientific. For the Cu-base TLP bonding, four alternating layers of Ti and Cu were stacked together to form the TLP interlayer, achieving a total thickness similar to the Nb foil and an overall composition of Ti-56.6 at.% Cu. For the Nb-base TLP bonding, a single layer of Nb foil was used as the TLP interlayer. The TLP assemblies were subsequently placed in a RD Webb vacuum furnace and evacuated to high vacuum ($10^{-6}$ torr) prior to heat treatment. The Cu-base bonding assembly was heat-treated at 1120 °C for 36 h and the Nb-base assembly was heat-treated at 1200 °C for 16 h; for both chemistries, there was no applied external load, the heating rate was 10 ºC/min and the cooling rate was ~ 10 ºC/min until 600 ºC followed by furnace cooling to room temperature.

For both joints, the compositions with the highest solute concentrations at the joint interface (i.e., Cu or Nb) were confirmed using EDS and reference alloys were reproduced via arc melting (Arc Melter 0.5, Edmund Bühler GmbH) to evaluate the transformation behavior of the joint composition via DSC.

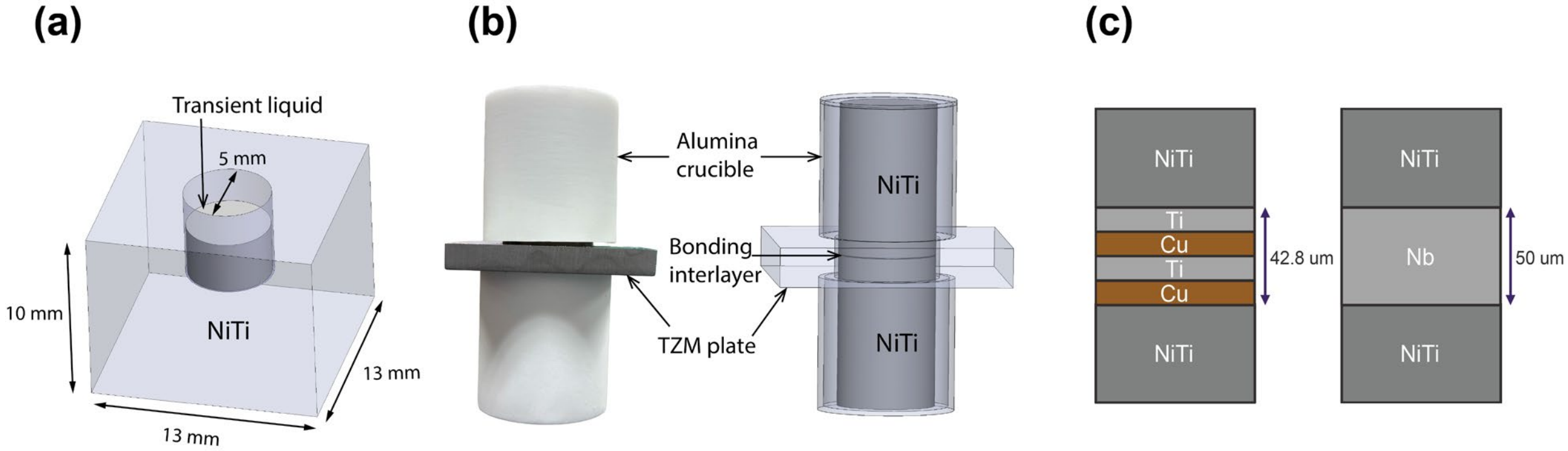

*Figure 1. Illustrations and images of the geometries used in the solidification and TLP bonding experiments. (a) Schematic of the setup used for quantifying the kinetics and thermodynamics for both TLP bonding systems. TLP precursors were added in powder form factors into the machine wells in (a) and subsequently heated in a controlled environment. (b) Fixture to carry out TLP bonding of the NiTi rods, which included a TZM plate for alignment and alumina crucibles. (c) Schematic of the interlayer foil geometry for the two TLP bonding systems in (b).*

## 2.3. Microstructural Characterization

To examine the microstructures after the heat treatments, cross-sections were mechanically ground and polished down to 50 nm colloidal silica mixed with 5 vol.% $H_2O_2$. Scanning electron microscopy (SEM, FEI Quanta 650 SEM) was performed on the NiTi cubes and joint interfaces. Energy dispersive spectroscopy (EDS) was used to analyze chemical compositions and electron backscattered diffraction (EBSD) was used to characterize the crystallographic structures and orientations. The solidification distance in the NiTi substrates was determined using EDS line scans and was defined as the distance from the solid–liquid interface to the point where the Cu or Nb concentration dropped to 0.5 at.%.

## 2.4. Thermal and Mechanical Testing

Differential scanning calorimetry (DSC) was performed on the as-received NiTi and arc-melted NiTiCu and NiTiNb samples using a DSC250 system (TA Instruments). Tests were conducted under a nitrogen atmosphere with a cooling/heating rate of 10 K/min. Phase transformation temperatures were determined from the resulting thermograms.

Tensile specimens were prepared via wire EDM as a scaled-down ASTM E8/E8M, with a gauge length of 23 mm and the other relevant dimensions scaled accordingly to keep the same geometric ratios [40]. Tensile tests of the joints, arc-melted samples and as-received NiTi were performed using an MTS Criterion load frame at room temperature with a crosshead speed of 0.006 mm/s, corresponding to a nominal strain rate of order $10^{-4}$ $sec^{-1}$. For cyclic tensile testing, the dog bone-shaped specimens were loaded from 25 MPa to 425 MPa for 10 cycles before pulling to failure.

In-plane strain mapping was performed using digital image correlation (DIC). A random speckle pattern was applied to the sample surface using an airbrush. Images were acquired at an acquisition rate of 2 Hz using a 4.2-megapixel camera with a 1x lens. Strains were calculated using VIC-2D. Virtual extensometers were used to extract the nominal strain across the gauge length, and averaged strains in distinct regions (e.g., base NiTi or joint) were extracted to quantify the individual material behavior.

To further probe the localized NiTi-joint mechanical response, nanoindentation was conducted with a KLA Instruments iMicro nanoindentation tester equipped with a Berkovich tip, using a 140 × 5 grid with 50 μm spacing and a 75 mN indentation force. Hardness and modulus values were calculated using the Oliver-Pharr method and were averaged along the width direction and plotted as a function of distance along the length of the dog bone.

# 3. Results and Discussion

## 3.1. CALPHAD Evaluation of TLP Chemistries

While TLP bonding proceeds through four dominant stages (melting, widening, solidification, homogenization), isothermal solidification is 3-4 orders of magnitude slower than the liquid-based interlayer dissolution and widening processes [41]. The rate limiting kinetics of isothermal solidification are governed by solid-state diffusion of the solute element X into the substrate. The relationship between the liquid interlayer thickness and the isothermal solidification time for a binary system can be approximated [38] using Fick's 2nd law and applying a Stefan condition for mass balance:

$$t_3 = \frac{W^2}{4K^2 D_s}, \quad (1)$$

where $D_s$ is the solid diffusion coefficient of the solute in the base metal, $W$ is the liquid interlayer thickness before isothermal solidification, and the constant $K$ can be numerically solved:

$$\sqrt{\pi} \cdot K[\text{erf}(K) + 1] \cdot e^{-K^2} = \Omega$$
$$\Omega \stackrel{\text{def}}{=} \frac{C^{\alpha L} - C^{\alpha 0}}{C^{L\alpha} - C^{\alpha L}} \quad (2)$$

Here, $C^{L\alpha}$ is the solute concentration in the transient liquid, $C^{\alpha L}$ is the solute concentration in the solid (in equilibrium with the transient liquid), and $C^{\alpha 0}$ is the solute concentration in the base metal. Equation 2 is transcendental and therefore must be solved numerically for $K$.

Since the liquid interlayer thickness can be tuned independently of the elemental system, a figure of merit (FoM) can be defined for TLP:

$$FoM = \frac{1}{K^2 D_s} \quad (3)$$

The $FoM \propto t_3$ scales directly with the isothermal solidification time, so a lower $FoM$ is preferred. This relationship is derived from the analytical solution for a binary system, Equation (1), and it warrants emphasis that it is not possible to assign a single $FoM$ to a TLP system containing more than two elements. Unlike a binary, $C^{L\alpha}$ and $C^{\alpha L}$ do not necessarily remain constant during isothermal solidification and the interface composition will shift depending on the relative diffusivity of each element (now defined as $D_x$ to be more general) in the substrate and their compositional dependence. Numerical modeling is needed to calculate the true solidification time, but this is a non-trivial effort and diffusion modules (like DICTRA) struggle to produce consistent results. For non-binary systems, the variables in Equations (1-3) therefore take upper and lower bounds spanning the relevant tie-lines; these bounds are presented in Table S1.

For this study, the $FoM$ for the third element "X" was used as a heuristic estimate for the true relative solidification time. Table 1 shows the minimum and maximum $FoM$ values (associated with every relevant tieline along the solid-liquid region) for each system at the lowest temperature at which a liquid was present. When $\Omega < -1$ there is no value of $K$ that satisfies Equation (2); physically a negative $\Omega$ value implies the interface will move in the opposite direction, driving melting rather than solidification. As a result, these systems have no valid $FoM$ values and were not included in Table 1. For systems that have both positive and negative $\Omega$ values, such as Cu, $FoM_{min}$ values that correspond to the lowest, positive $\Omega$ are reported.

*Table 1 TLP FoM calculations using Equation (3) and the lower/upper values for $K_x$ and $D_x$ in Table S1.*

| Ternary Element | Temp (K) | | |
|---|---|---|---|
| Ta | 900 | $1.86 \times 10^{17}$ | $4.09 \times 10^{18}$ |
| Mn | 1100 | $9.33 \times 10^{14}$ | $3.27 \times 10^{15}$ |
| Hf | 1200 | $1.07 \times 10^{13}$ | $4.11 \times 10^{15}$ |
| Nb | 1200 | $4.32 \times 10^{13}$ | $8.44 \times 10^{13}$ |
| Zn | 1200 | $2.54 \times 10^{13}$ | $3.93 \times 10^{13}$ |
| Cu | 1300 | $7.59 \times 10^{11}$ | $7.17 \times 10^{13}$ |
| W | 1300 | $2.30 \times 10^{17}$ | $3.42 \times 10^{19}$ |
| Zr | 1300 | $8.03 \times 10^{11}$ | $1.17 \times 10^{12}$ |

The eight down-selected elements in Table 1 are all viable TLP systems for NiTi. Notably, many of the refractory elements (Hf, Ta, W) have chemical behavior analogous to that of Nb, where a low-temperature liquid appears in their Ni-Ti-X ternaries, and two common substitutional elements in NiTi (Cu and Zr) were identified. The four ternary elements (Nb, Zn, Cu, Zr) highlighted in Table 1 possess the lowest *FoM* values by at least an order of magnitude, and thus are estimated to be the most promising. Zn was excluded as a possible TLP system due to its high vapor pressure; Zr, despite possessing the lowest *FoM* values, was excluded due to its high oxygen affinity. Both Cu and Nb were selected for a detailed study due to their similar *FoM* values and contrasting roles on the NiTi sublattice: Cu substitutes for Ni, whereas Nb substitutes for Ti.

## 3.2. Thermodynamic equilibrium and kinetics during isothermal solidification

To evaluate whether TLP bonding will be active in these two ternary systems, isothermal holds were carried out for pre-determined times and temperatures using the specimen geometries outlined in Figure 1(a). The Cu-base system was heat treated at 1120 °C and the Nb-base system was heat treated at 1200 °C. These temperatures were chosen for their respective systems to ensure that any compositional fluctuations (within ±5 at.% of the nominal interlayer composition) fall entirely within the liquid phase region. By tracking the solid/liquid interface position versus time, the kinetics and rate-limiting behavior of the Cu-base and Nb-base systems can be quantified and compared.

The solidification distances for the Cu-base and Nb-base systems are shown in Figure 2. The long hold times required to separate successive interface positions beyond their measurement uncertainty restricted the full time series to a single temperature for each ternary; two isochronal sets, each spanning three temperatures, were collected to extract activation energies for both chemistries. These datasets were sufficient to evaluate whether an isothermal solidification mechanism was active due to the diffusion-limited kinetics that govern TLP bonding. Following the discussion above, the isothermal solidification distance $d$ can be approximated as:

$$d = K\sqrt{4D_x t}, \quad (4)$$

where the terms were defined above in Equations (1) and (2). Importantly, the solidification distance is expected to follow a $t^{0.5}$ relationship – provided the interface composition and bulk diffusivity do not appreciably change during solidification.

The dashed lines in Figure 2(a&b) are fits of the isothermal data to Equation (4); the agreement indicates that solidification is governed by bulk diffusion. At similar temperatures and equal times, the solidification distance of the Cu-base system is ~25% larger than that of the Nb-base system; at their respective minimum bonding temperatures (1120 °C for Cu-base, 1200 °C for Nb-base), the distances become comparable (Figure 2a,b). This ordering follows the solutes themselves: solute diffusivity in a given matrix correlates inversely with the solute melting temperature, and Nb, with a far higher melting point than Cu, diffuses more sluggishly through NiTi at a given temperature. The diffusivity can be examined more closely by plotting ln $d$ against 1/T and extracting the activation energy $E_A$. Figure 2(c)-(d) shows Arrhenius plots for both TLP bonding systems. The activation energy values were determined for each system and are included in Table 2.

Analogous to the solidification data, the activation energy of NiTi-Nb interdiffusion is larger than that of NiTi-TiCu. In addition, for each system, the activation energy values determined at different heat treatment times are in good agreement. It has been previously reported that the liquid composition can shift during isothermal solidification when more than two elements are present, which could result in a variable activation energy [39]. However, this shift occurs when the liquid volume is small compared to the solid, as in TLP bonding, whereas the liquid and solid phase volumes in these experiments are of the same order of magnitude. It was hypothesized in the same study [39] that an “equilibrium liquid composition” could be achieved for large liquid volumes and long solidification times. The relatively constant activation energy across these two timescales suggests this equilibrium may have been achieved.

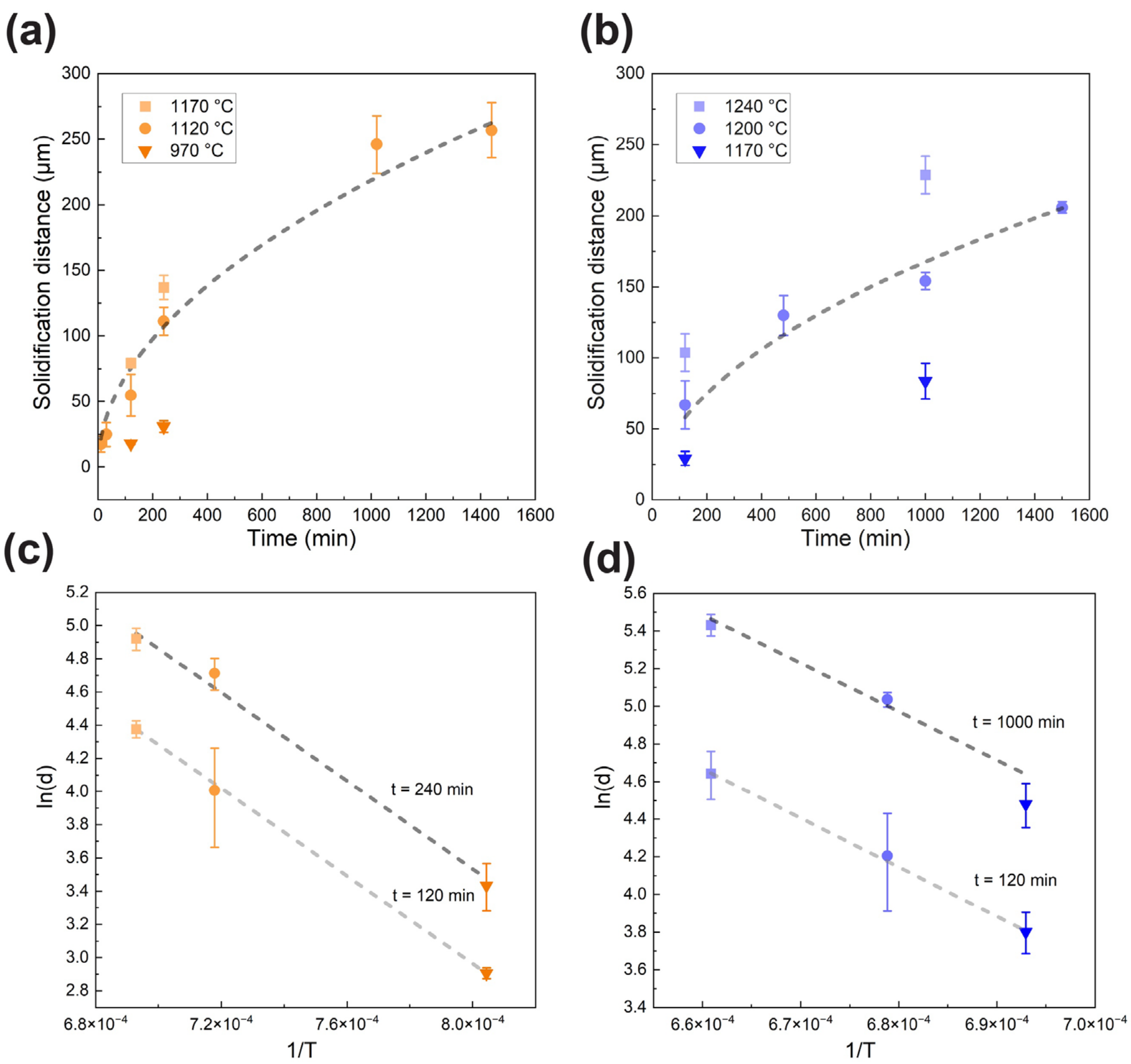

*Figure 2 Comparison of solidification data between (orange) Cu-base and (blue) Nb-base TLP systems. (a, b) Solidification distance versus time for the Cu-base and Nb-base compositions across three different temperatures. (gray dashed lines) $t^{1/2}$ fit to solidification data at a single temperature (circles). (c, d) Arrhenius plots of the solidification distance versus 1/T for two isochronal sets.*

*Table 2. Activation energies for the Cu-base and Nb-base systems from Arrhenius plots in Figure 2(c,d).*

| | | |
|---|---|---|
| Cu-base | 2.18 ± 0.04 eV (120 min) | 2.00 ± 0.24 eV (240 min) |
| Nb-base | 4.28 ± 0.94 eV (120 min) | 4.36 ± 0.11 eV (1000 min) |

## 3.3. Microstructural evolution in the joint region during TLP bonding

The microstructure of the as-received NiTi is shown in Figure 3 and consists entirely of the B2 austenite phase. The grain morphology is equiaxed, with a grain diameter of 26.9 ± 14.3 μm and a small fraction (<10%) of low angle grain boundaries. A $TiC_xO_{1-x}$ impurity phase is present at a low fraction (~0.13%), which forms from Ti gettering interstitial impurities [42-44]. During EBSD indexing, this phase was identified as cubic $Ti_2C$ with space group $Fd\bar{3}m$. However, EDS analysis identified the composition to be $TiC_xO_{1-x}$, which has a rock-salt structure. This phase was also frequently misidentified as B2 NiTi so its actual fraction may exceed the value reported in Table 3. $Ti_2Ni/Ti_4Ni_2O$ precipitates were also identified, but their phase fraction was negligible.

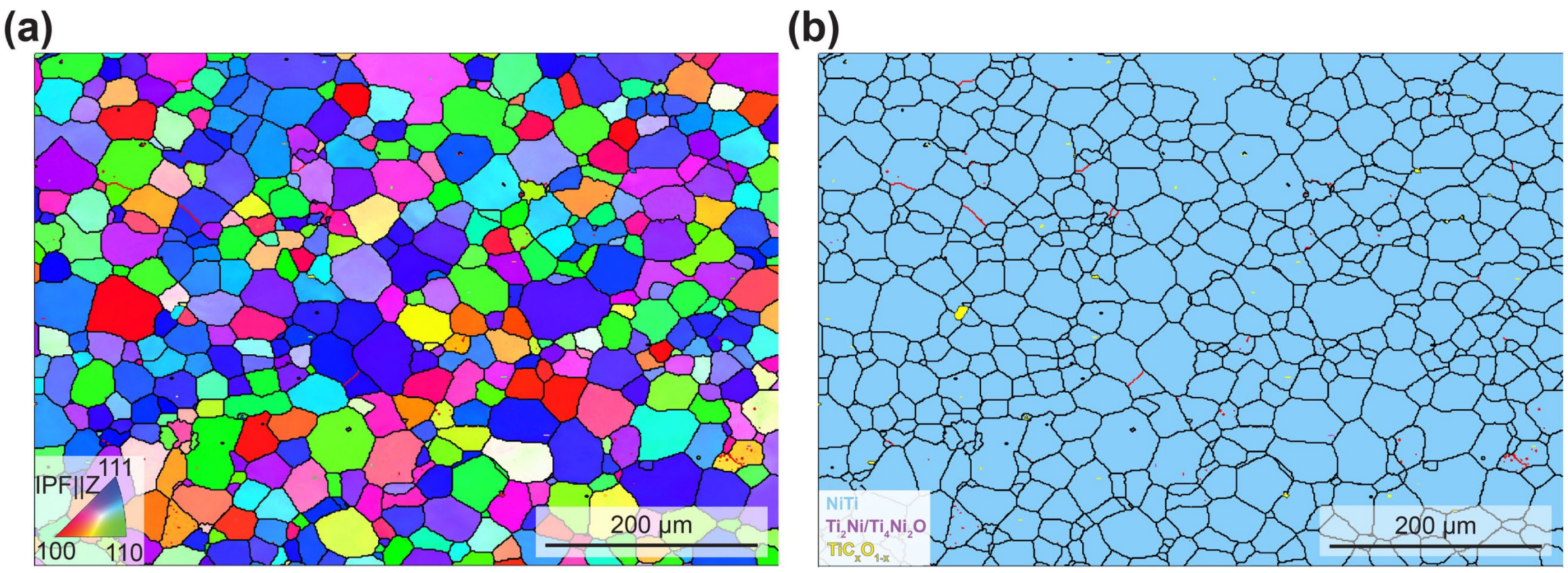

*Figure 3 Microstructure of the as-received NiTi material. (a) Inverse pole figure IPF-Z map and (b) phase map obtained from EBSD. Low-angle grain boundaries (2–5°) are indicated by red lines, and high-angle boundaries by black lines.*

Representative electron micrographs of both bonded specimens, Figure 4(a) and Figure 5(a), show fully dense joint regions, indicating that the transient liquid was able to effectively wet both substrates and no delamination occurred during bonding and cooling. Dark particles are observed in both sets of electron micrographs and correspond to the $TiC_xO_{1-x}$ phase that was identified in the as-received material. Importantly, these interstitial compounds are uniformly distributed throughout the material and are not confined to the joint region; chemical analysis of the material after the bonding heat treatment revealed that the overall oxygen content increased from 270 ppm to 560 ppm and 370 ppm for the Cu and Nb joints, respectively. The particle diameters are larger than those observed in Figure 3, which is likely due to Ostwald ripening during the bonding heat treatments. An additional phase (dark gray) can be observed in the electron micrographs and is attributed to $Ti_2Ni/Ti_4Ni_2O$; this phase will be discussed below.

The low-magnification IPF-Z maps in Figure 4(c) and Figure 5(c) show large grains, with average values of 213 ± 202 µm for the Cu-base joint and 146 ± 126 µm for the Nb-base joint; the coarser, potentially bimodal grain structure observed in the Cu-base system may be due to the longer bonding duration (36 hr at 1120 °C) when compared to the Nb-base system (16 hours at 1200 °C). However, due to the limited field of view, the number of grains in the micrographs is low and thus the statistics may not be truly representative. Both joint regions exhibit the characteristic grain impingement that arises from completion of isothermal solidification. If residual liquid were present during cooling after bonding, small grains would have nucleated in this region and be clearly visible in the micrograph. EDS line scans across the joint interfaces, Figure 4(b) and Figure 5(b), confirm that the Cu and Nb solutes have diffused out of the joint region and diluted into the NiTi substrates, with peak concentrations of Cu (1.9 at.%) and Nb (4.3 at.%) where the ~50 μm-thick foil interlayers were initially located. These solute concentrations are below their respective solid solution limits in NiTi and provide further evidence that isothermal solidification proceeded to completion [32, 45].

*Table 3 Phase fractions (vol.%) within ±50 μm from the joint interface for NiTi–TiCu–NiTi and NiTi–Nb–NiTi joints, determined from EBSD phase identification*

| | NiTi B2 (vol. %) | $Ti_2Ni/Ti_4Ni_2O$ (vol. %) | $TiC_xO_{1-x}$ (vol. %) | Low-angle grain boundary (2-5°) fraction |
|---|---|---|---|---|
| As-received NiTi | - | - | 0.13 | 8.63 |
| NiTi joined with TiCu | 99.51 | 0.44 | 0.06 | 5.28 |
| NiTi joined with Nb | 98.79 | 1.18 | 0.02 | 5.67 |

The extent to which isothermal solidification has occurred can be more rigorously evaluated by examining the phase fractions in the joint interface region (Table 3). For both ternary systems, the B2 phase dominates (~99%) with small fractions of other phases present. In contrast to the as-received material, where this phase was negligible, the $Ti_2Ni/Ti_4Ni_2O$ phase appears at measurable fractions and sparsely decorates the joint interface, Figure 4(d) and Figure 5(d); the compositions of the non-NiTi intermetallic phases in both joints are included in Table 4. It is possible these phases formed due to the presence of small pockets of unsolidified liquid after bonding or precipitated during the cooling process [32]. The fraction of these phases is slightly higher in the Nb-base system – suggesting the Nb-base bond had more unsolidified liquid and/or the driving force for the second phase is larger in this ternary system. However, the overall fraction of these impurities is relatively low in both interlayer chemistries and is expected to have a negligible impact on macroscopic properties.

*Table 4 Compositions (at.%) of identified secondary phases*

| | Ti | Ni | Cu | Nb | C | O |
|---|---|---|---|---|---|---|
| $Ti_2Ni/Ti_4Ni_2O$ (Cu-base bond) | 56.6 | 25.6 | 4.5 | - | - | 13.3 |
| $Ti_2Ni/Ti_4Ni_2O$ (Nb-base bond) | 54.3 | 29.1 | - | 4.0 | - | 12.6 |
| $TiC_xO_{1-x}$ | 45.7 | 9.9 | - | - | 30.4 | 13.8 |

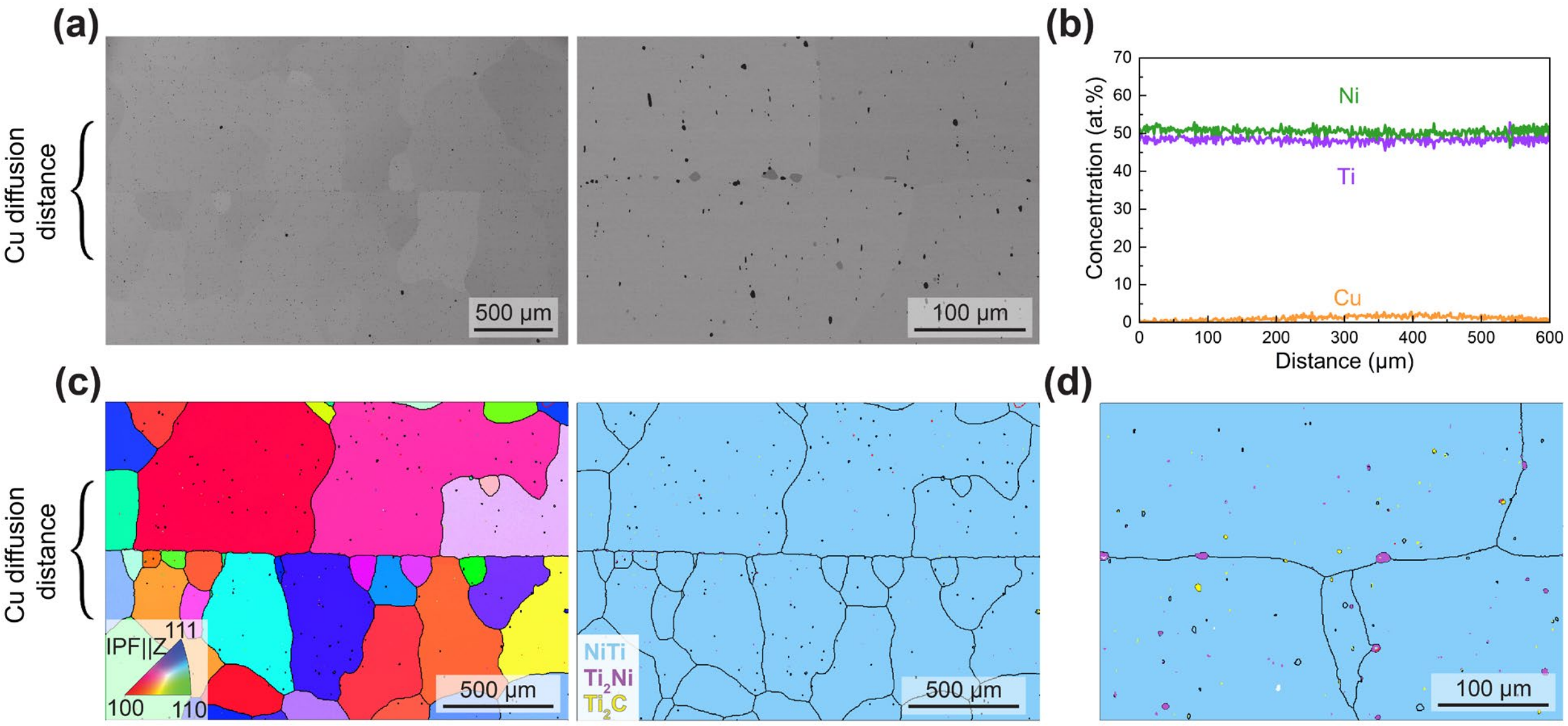

*Figure 4 Representative microstructures after Cu-base TLP bonding at 1120 °C for 36 hours. (a) Secondary electron micrographs of the joint and adjacent area. (b) EDS line scan passing through the joint to quantify the residual Cu content in the joint region. (c) Inverse pole figure (IPF-Z) map and phase map of NiTi-Cu joint cross-section obtained from EBSD, revealing a horizontal line due to grains from the two substrates impinging during isothermal solidification. (d) Higher-magnification EBSD phase map of the joint region, showing individual phases: (blue) austenite; (purple) intermetallic; (yellow) carbide; and (smaller circles) unidentified oxides. Low-angle grain boundaries (2–5°) are indicated by red lines, and high-angle boundaries by black lines.*

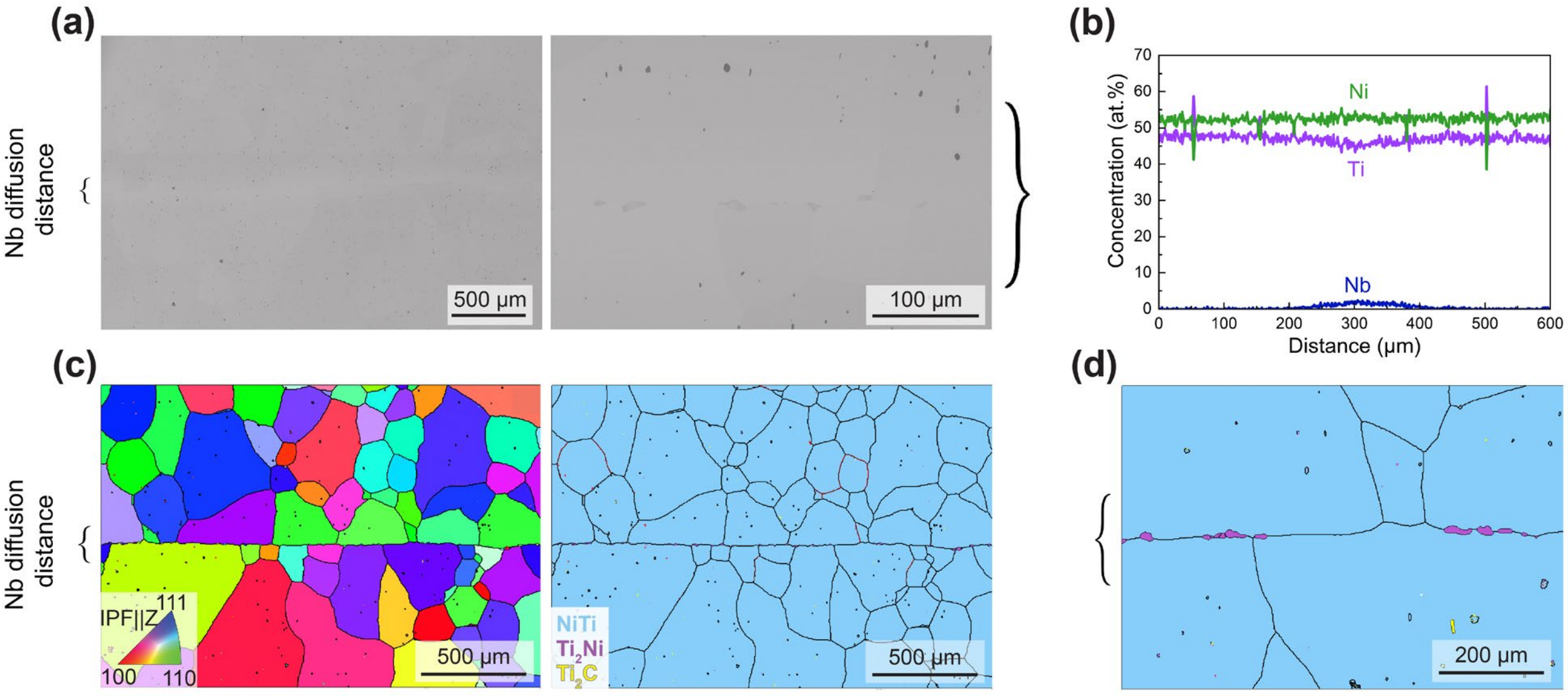

*Figure 5 Representative microstructures after Nb-base TLP bonding at 1200 °C for 16 hours. (a) Secondary electron micrographs of the joint and adjacent area. (b) EDS line scan passing through the joint to quantify the residual Nb content in the joint region. (c) Inverse pole figure (IPF-Z) map and phase map of NiTi-Nb joint cross-section obtained from EBSD, revealing a horizontal line due to grains from the two substrates impinging during bonding. (d) Higher-magnification EBSD phase map of the joint region, showing the individual phases: (blue) austenite; (purple) intermetallic; (yellow) carbide; and (smaller circles) unidentified oxides. Low-angle grain boundaries (2–5°) are indicated by red lines, and high-angle boundaries by black lines.*

To better estimate the role of the dilute solutes on the properties of NiTi, bulk specimens were prepared via arc melting using the maximum Cu and Nb concentrations measured across the joint interface. The DSC results of the arc-melted samples, along with the as-received material, are presented in Figure 6. The as-received NiTi has an austenite start temperature ($A_s$) of ~ -25 ºC and an austenite finish temperature ($A_f$) of ~1.4 ºC, bracketing the nominal -15 °C transformation temperature provided by the manufacturer. Compared with the as-received NiTi, the NiTi-Cu alloy has slightly higher transformation temperatures but remains austenitic above room temperature. This composition also has a reduced hysteresis, consistent with previous reports on NiTi-Cu SMAs [46]. The DSC curve for the NiTi-Nb alloy has no detectable phase transformation peaks from -100 to 140 ºC. This lack of observable peaks could be attributed to a martensite finish temperature ($M_f$) below -90 °C or suppression of the phase transformation behavior at this particular alloy composition. However, the former interpretation is more likely because Nb is known to reduce the martensite start temperature in NiTi, with one study reporting an $M_f$ of approximately -156 °C [47].

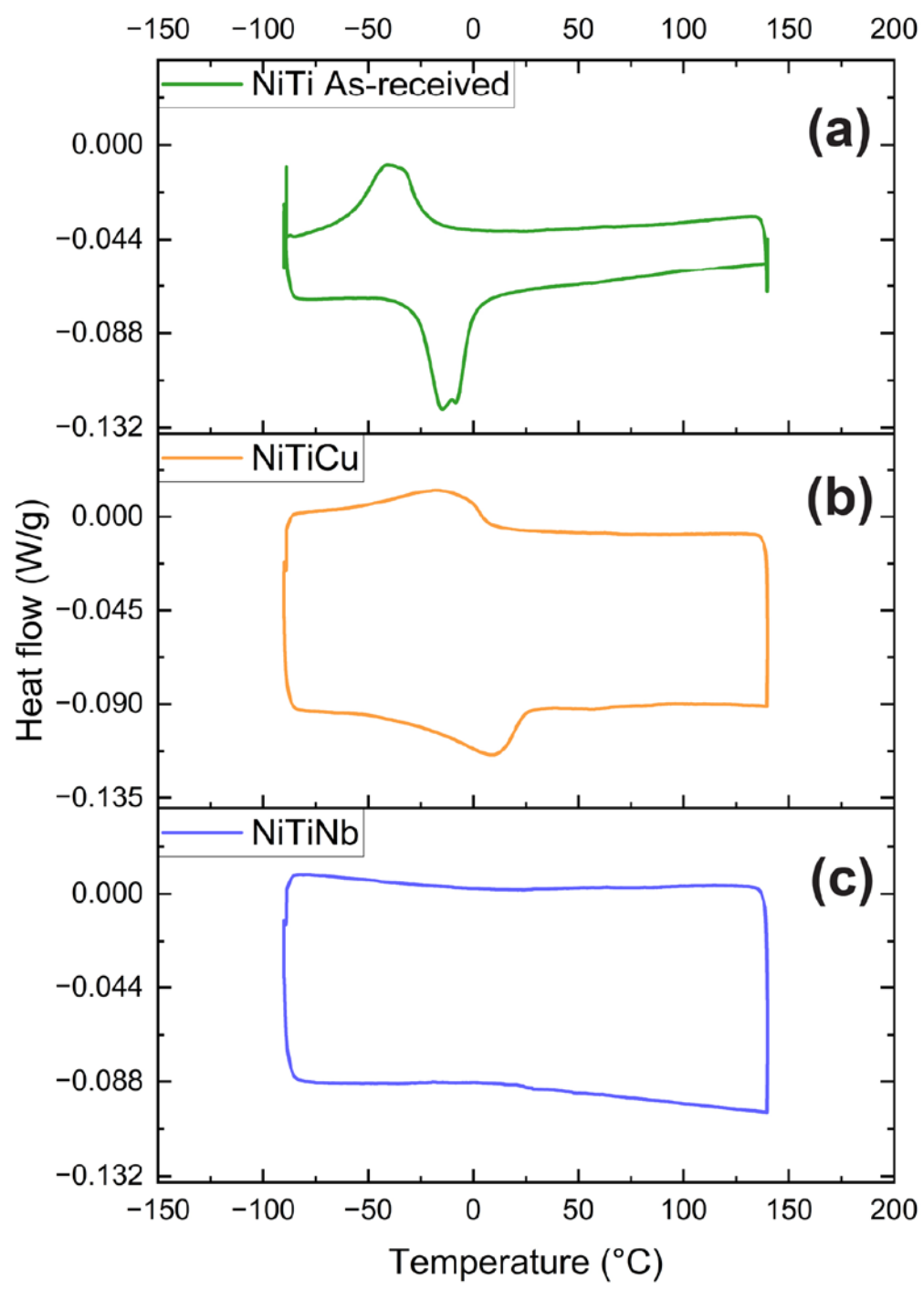

*Figure 6 Differential scanning calorimetry comparison of the: (a) as-received NiTi, and (b,c) arc-melted, bulk NiTi-Cu and NiTi-Nb specimens, respectively. The Cu and Nb compositions were selected to match the largest Cu and Nb fractions measured within the joint regions in Figures 4(b) and 5(b). The as-received material is austenitic, with a transformation temperature below RT. For NiTi-Cu, Cu substitutes for Ni on the B2 lattice and raises the transition temperature. For NiTi-Nb, Nb is expected to substitute for Ti and lower the transition temperature. It is possible that this material no longer exhibits this phase transformation.*

## 3.4. Mechanical behavior of the joints

### 3.4.1. Macroscopic behavior of the two joint chemistries

Stress-strain curves from uniaxial tensile pull-to-failure tests are shown in Figure 7(a) for the as-received NiTi (baseline) and the as-received NiTi that underwent a 1120 °C, 36 hr heat treatment to mimic bonding conditions. Compared to the as-received material, the heat-treated NiTi has a lower martensite onset stress ($\sigma_M^s$), decreasing from 645 MPa to 401 MPa, while the ultimate tensile strength (UTS) slightly decreased from 922 MPa to 868 MPa. This reduction in $\sigma_M^s$ and UTS is attributed to the removal of cold work and grain coarsening during the extended high temperature exposure. Additionally, the ductility decreased substantially from 22.6% to 6.8%, which could be due to strain incompatibility among the coarsened grains or an increased interstitial impurity content introduced during the heat treatment; similar to the bonded specimens, the oxygen content after heat treatment increased from 270 to 560 ppm.

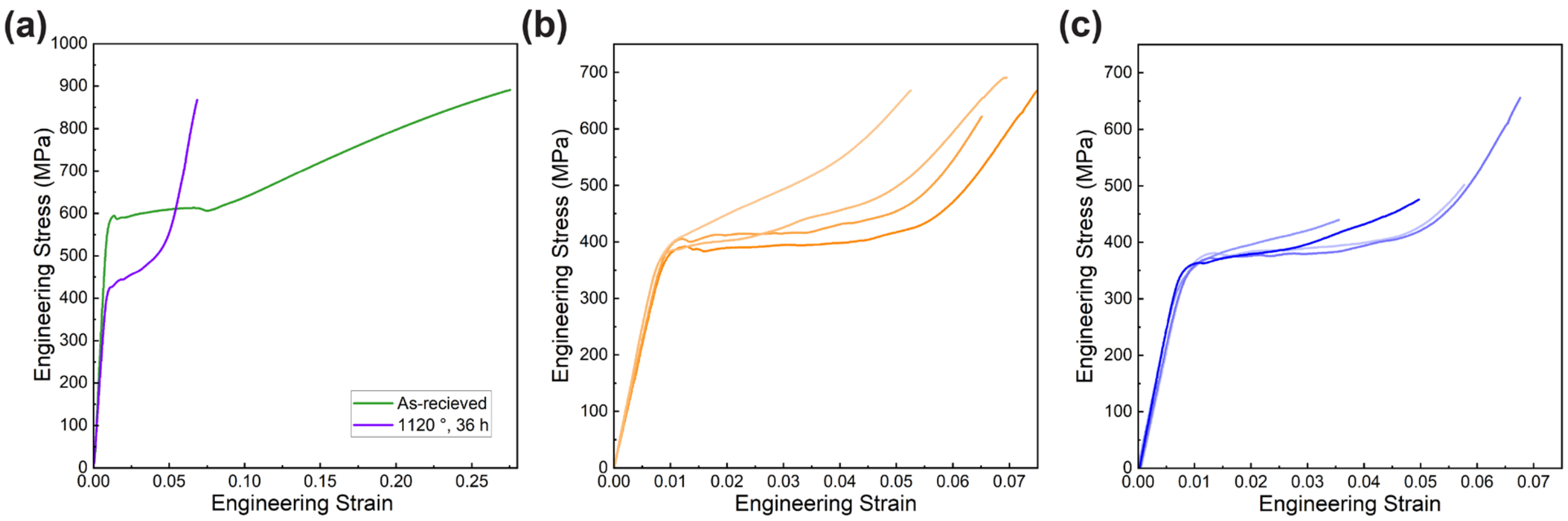

*Figure 7 Tensile stress-strain curves. (a) As-received NiTi (green) and as-received NiTi heat treated at 1120 °C for 36 h (purple). (b) Quasistatic curves for the Cu-base TLP joints. (c) Quasistatic curves for the Nb-base TLP joints.*

Figure 7(b,c) show the stress-strain response for the Cu-base and Nb-base bonds pulled to failure. The Cu-base bond has an average $\sigma_M^s$ of 375 ± 12 MPa, martensite finish stress ($\sigma_M^f$) of 488 ± 75 MPa, and UTS of 663 ± 29 MPa. The Nb-base bond has an average $\sigma_M^s$ of 356 ± 18 MPa, $\sigma_M^f$ of 414 ± 5 MPa and UTS of 518 ± 95 MPa.

The estimated joint efficiencies – comparing the stresses achieved by the joined specimens to those by the heat-treated as-received NiTi – are summarized in Table 5. In terms of $\sigma_M^s$ and $\sigma_M^f$, both efficiencies are greater than 75%. However, the Cu-base bonds exhibit higher transformation stresses, with efficiencies exceeding 90%. While the UTS joint efficiency of the Nb-base bond (59.6%) is lower than that of the Cu-base bond (76.2%), both are higher than values reported for comparable NiTi joining methods, such as solid-state diffusion bonding [48] and brazing [49]; the UTS of the Cu-base bond is higher than reports on joints fabricated via NiTi friction stir welding [26].

*Table 5 Joint efficiencies of both interlayer chemistries determined by normalizing the macroscopic property ($\sigma_M^s$, $\sigma_M^f$, UTS) of the joined specimen by the corresponding value from the heat-treated as-received NiTi in Figure 7(a).*
** Only 2 out of 4 tested samples reached the $\sigma_M^f$ without failure*

| | Cu-base bond | Nb-base bond |
|---|---|---|
| Joint efficiency [calculated using $\sigma_M^s$] | 93.6% | 88.8% |
| Joint efficiency [calculated using $\sigma_M^f$] | 90.3% | 76.8%* |
| Joint efficiency [calculated using UTS] | 76.2% | 59.6% |

The cyclic stress–strain curves of the Cu-base and Nb-base bonds, Figure 8, exhibit the characteristic flag-shaped hysteresis loops associated with the superelastic behavior of NiTi. Both joints retained approximately 4% recoverable strain across all 10 cycles. With cycling, the transformation plateau steepened while $\sigma_M^s$ decreased. The emergence of this behavior during repeated cycling is likely due to the formation of dislocation substructures that assist the nucleation of favorably oriented martensite variants, lowering $\sigma_M^s$, while also impeding variant growth and coalescence, steepening the plateau [1, 50-52]. Nevertheless, differences emerge between the two chemistries as cycling proceeds.

For the Cu-base bond, $\sigma_M^s$ decreased from 398.4 MPa to 325.1 MPa over the 10 cycles, with a gauge-average permanent strain of 0.35%. The Nb-base bond behaved similarly: $\sigma_M^s$ decreased from 384.4 MPa to 330.0 MPa, with a permanent strain of 0.30%. Reported values of permanent strain in NiTi vary widely, ranging from approximately 0.25% to over 7.5% after 10 cycles [51-55] under loading conditions comparable to those used in this study (i.e., comparable stress or strain amplitudes and test temperatures). Nevertheless, it can be concluded that TLP bonding has a limited impact on strain accumulation during cycling. To better resolve the differences between these two chemistries and gain further insight into their response to mechanical load, the spatially resolved strain was examined.

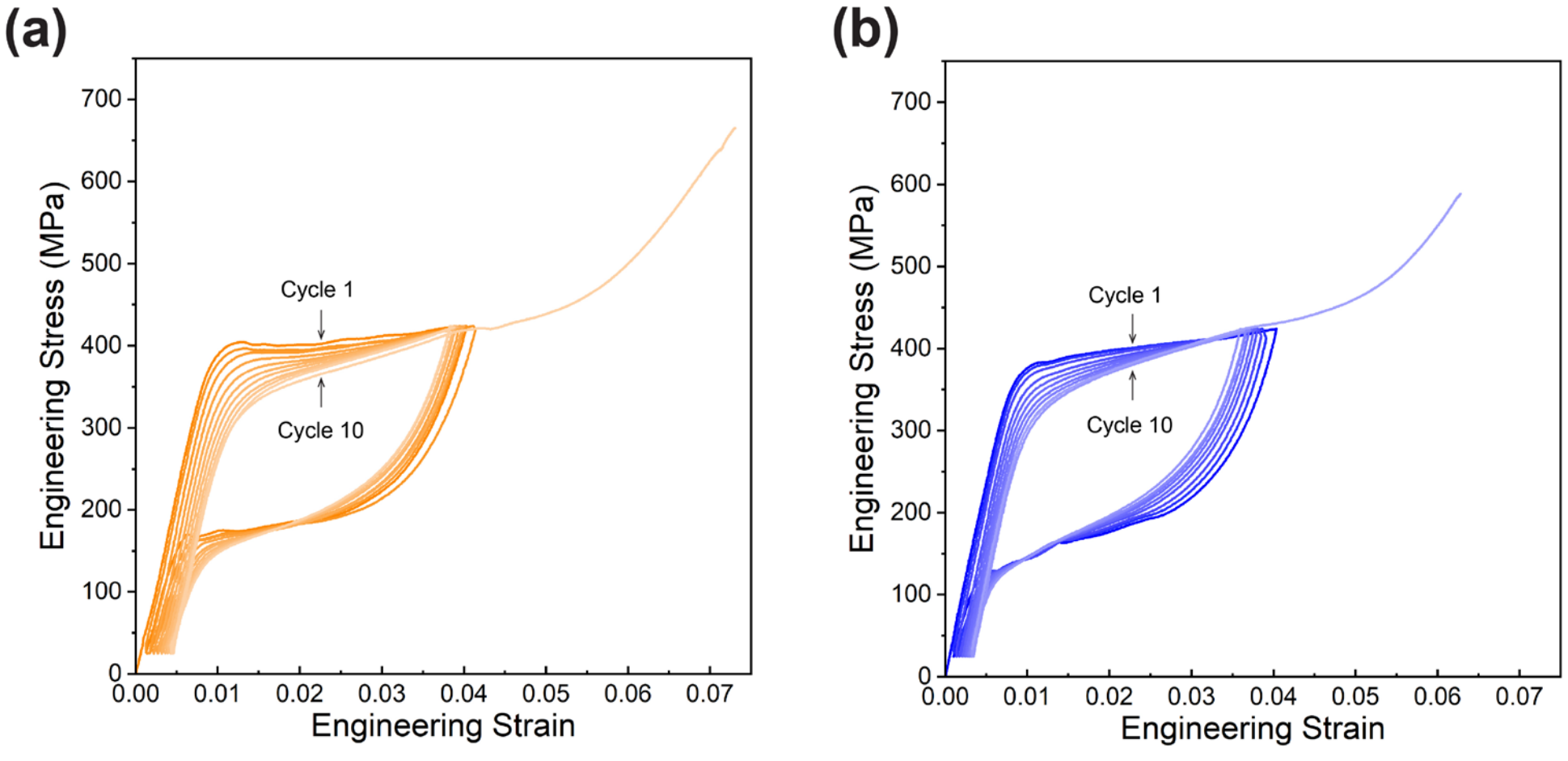

*Figure 8. Cyclic tensile testing of the TLP joints: 25–425 MPa loading for 10 cycles followed by monotonic loading to failure for the (a) Cu-base and (b) Nb-base bonds. Both joints survived all 10 cycles without failure.*

### 3.4.2. Examining the local deformation response

Figure 9 presents the strain distributions during the pull-to-failure tests, calculated via DIC. For clarity, three regions are defined: the left and right, which refer to the two NiTi substrates; and the joint, which refers to the diffusion-affected zone from the interlayer. For both chemistries, during initial loading, shear bands form on both sides of the joint, indicating effective load transfer. As deformation continues, the strain in the joint region lags the global strain, and the shear bands in the left and right regions grow towards the center of the gauge. This heterogeneous deformation continues until the left and right regions have fully transformed, and then the strain in the joint region increases – reaching a value comparable to, though still lower than, that of the left and right regions. With increased loading, the strain evolves more uniformly until failure. For the Cu-base chemistry, strain localization and eventual failure occur in the joint region for every sample, whereas the Nb-base specimens failed in the left or right regions

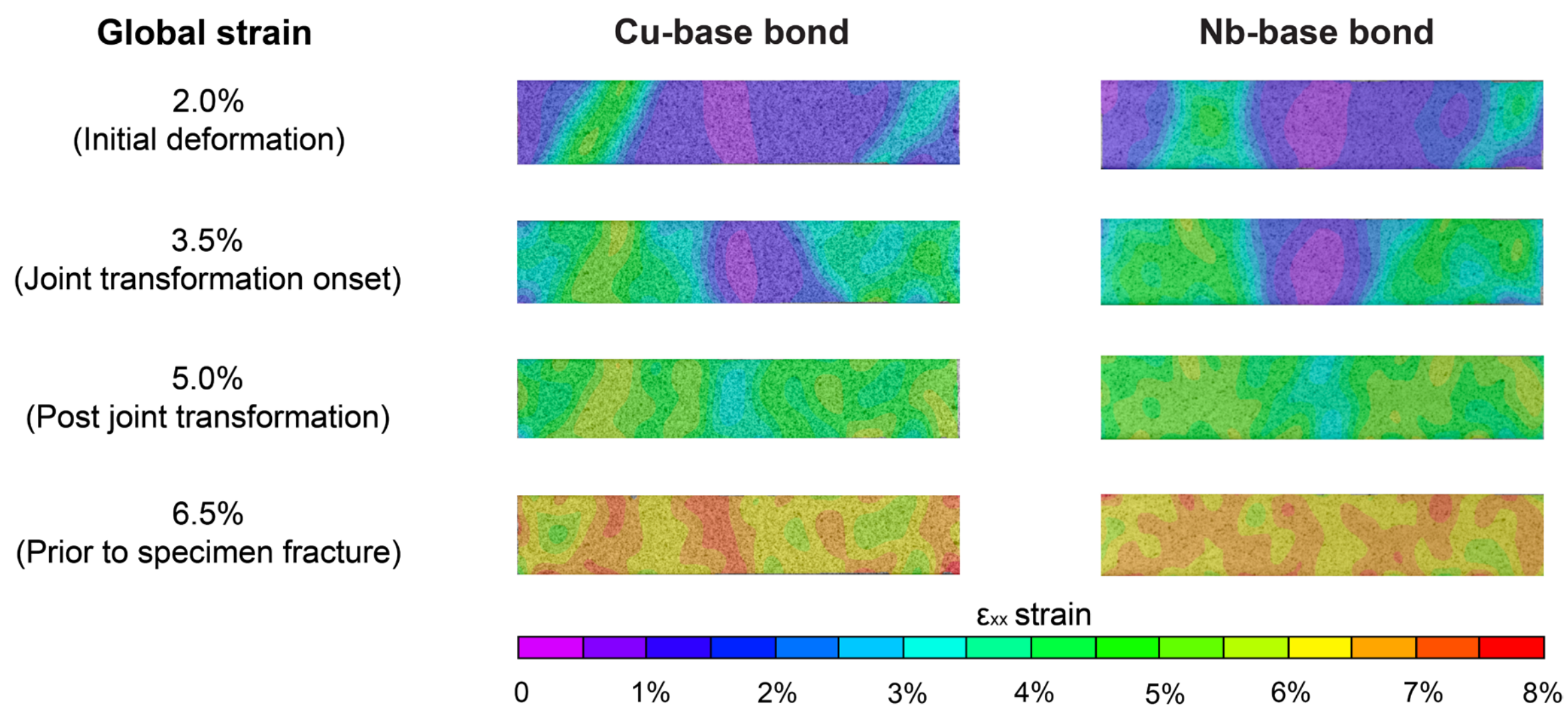

*Figure 9 Comparison of strain distribution during tensile pull-to-failure experiments. (a) Deformation initiates in the substrates, while the joint region undergoes limited deformation. (b) With continued loading, both the left and right regions continue to deform uniformly and relatively symmetrically, whereas the joint region lags. With increased loading, (c) the joint region eventually transforms, and (d) specimens finally fail having undergone relatively uniform deformation across the entire gauge length.*

Figure 10 presents the strain distributions during cyclic testing for both chemistries, calculated via DIC; the region-resolved metrics are summarized in Table S2. Consistent with the observations in Figure 9, the joint region has a lower strain during cycling, with averages on the first cycle of 1.12% and 1.45% for the Cu-base and Nb-base bonds, respectively. In contrast, the left and right regions reach higher peak strains, ranging from 4% to 5.2%. The two substrate halves do not reach identical peak strains: the difference is 0.49 to 0.52% for the Cu-base specimen and 0.55 to 0.67% for the Nb-base specimen over 10 cycles. This difference appears to be a geometric effect due to the bond being slightly off-center in the gauge region; the longer substrate section carries the higher strain (the right region for the Cu-base specimen and the left region for the Nb-base specimen, Table S2).

Similar to the pull-to-failure tests, shear bands associated with the martensite-to-austenite phase transformation nucleate in both the left and right regions and grow, advancing towards the joint region, Figure 10 (a&d). For both joint chemistries, the joint region shows limited strain and thus does not act as a strain concentrator during cycling. Despite cycling to the same peak stress, the Cu-base joint exhibits a lower strain than the Nb-base joint, Figure 10 (b&e), which is consistent with a higher stiffness in the Cu-base joint region. For both joint chemistries, the strain profiles evolve during cycling, with the load/unload curves becoming more symmetric with increasing cycle number – this is most apparent in examining the strain during unloading in Figure 10 (b&e). These trends indicate that the material is being trained during cycling, enabling the phase transformation to occur more easily. Similar behavior can be observed in the joint region, but the strain profiles in the Nb-base joint appear to become more symmetric than those in the Cu-base joint. Further evidence of training is seen in the accumulated strain versus cycle number, Figure 10 (c&f), which is growing at a decelerating rate with increasing cycle number.

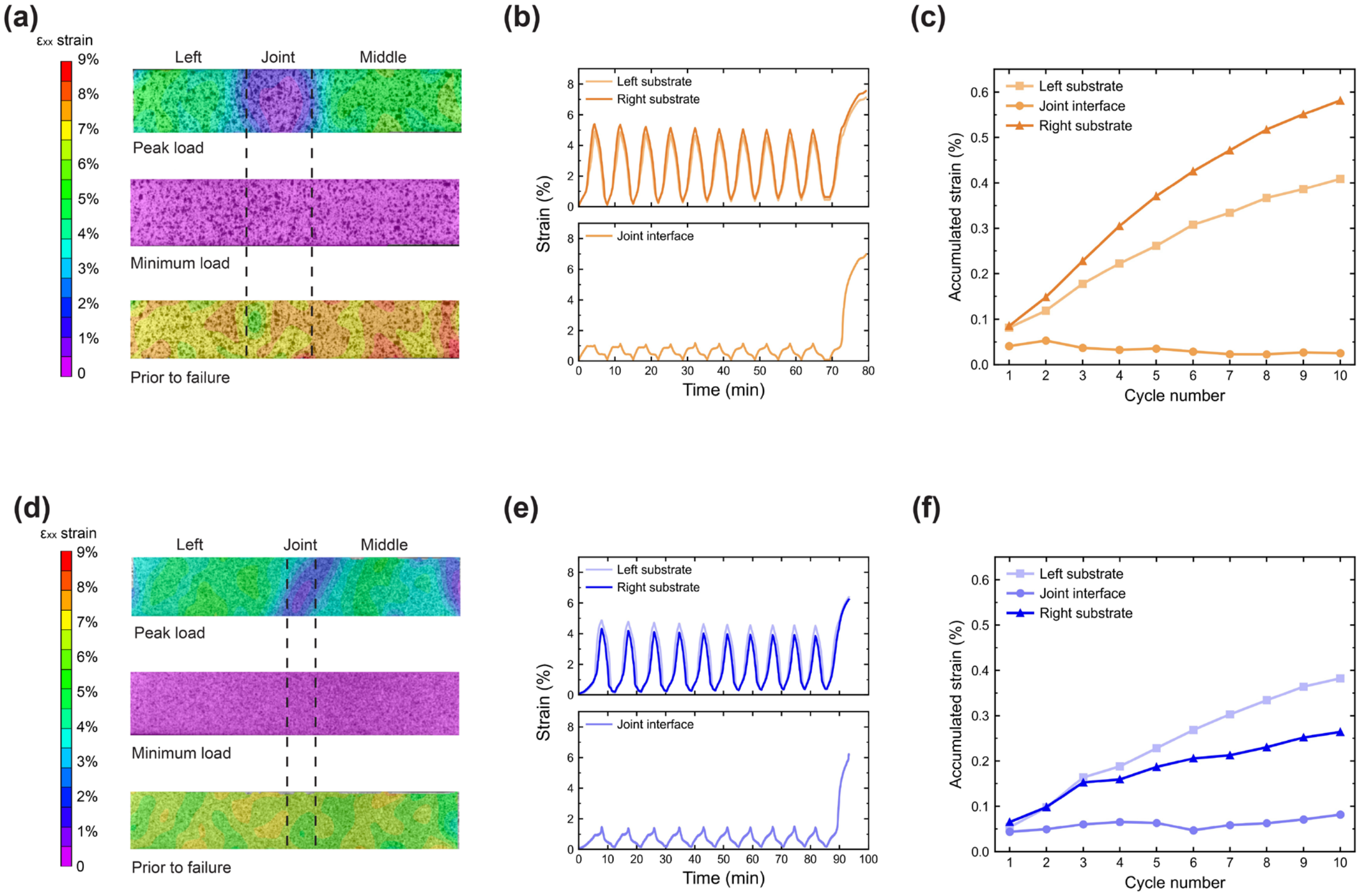

*Figure 10 Digital image correlation (DIC) mapping of the gauge region during cyclic testing of (a,b,c) Cu-base and (d,e,f) Nb-base TLP bonds. (a,d) $\varepsilon_{xx}$ strain maps during three representative loading conditions. (b,e) Local strain evolution over time extracted from ~700 μm wide regions in the left substrate, bond interface, and right substrate. (c,f) Accumulated strains at the end of each cycle for the three regions.*

The systematically lower strain within the joint region motivated local mechanical probing via nanoindentation, Figure 11. The results in Figure 11 show an increase in modulus within the joint regions for both ternary systems. The spatial footprints agree with the width of the joint regions identified in the DIC maps, ~4.5 mm for the Cu-base joint and ~1.5 mm for the Nb-base joint; however, these footprints exceed the solute ranges resolved by EDS, indicating that solute concentrations below the EDS detection limit still measurably raise the modulus. The modulus elevation corroborates the limited strain during tensile testing: the specimens are loaded in an isostress condition, and thus the local strain is inversely related to the local stiffness. The increased modulus is widest for the Cu-base joints, which explains the wide region of low strain in Figures 10 & 11. The increased hardness mirrors the modulus profile and provides an independent signature of solid-solution strengthening within the diffusion-affected zone.

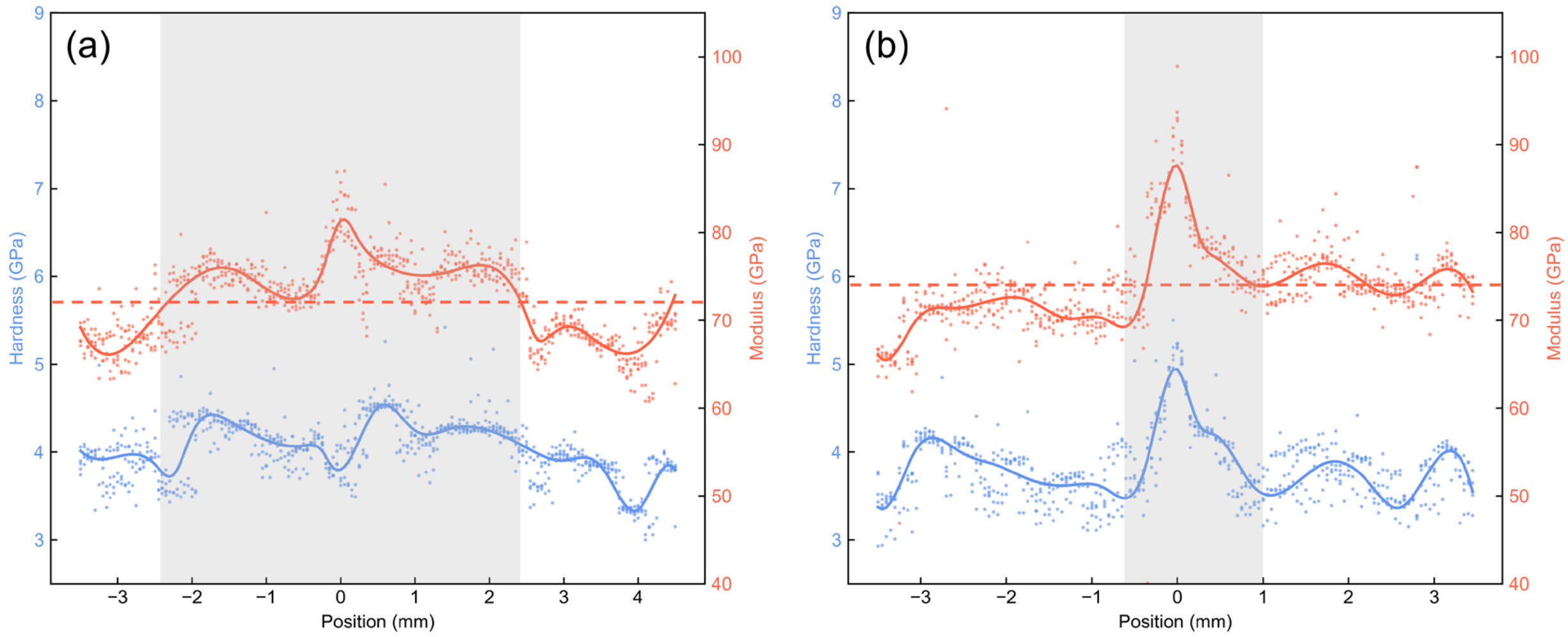

*Figure 11 Nanoindentation mapping of the joined specimens: (a) in the joint region for Cu; (b) in the joint region for Nb. Shaded gray boxes span the interface region and horizontal dashed lines indicate the average modulus value for the Cu (72 GPa) and Nb (74 GPa) bonds.*

### 3.4.3. Implications for joint design

For TLP joining of NiTi, the mechanical footprint of the joint is set by the solute diffusion field, which is controlled by the bonding time and temperature. Over the 10 cycles tested, solute additions from the TLP process produced joint regions that accumulated a factor of 7 (Cu-base) and 2 (Nb-base) less residual strain than the adjacent substrate regions (Table S2). The Nb-base joint region accumulated strain relatively monotonically, while the Cu-base joint showed substantially less accumulation; whether the low-accumulation advantage persists to fatigue-relevant cycle counts remains to be established. However, the substrate regions showed the opposite behavior, where the left and right regions of the Nb-base specimen accumulated lower strain than those of the Cu-base specimen.

This behavior is hypothesized to arise from both the mechanical properties and width of the joint region. The large diffusion distance of Cu is posited to help spread out the load, preventing joint localization, while the higher strength in the Nb-joint (using hardness as a surrogate) allows more efficient load transfer between the substrates. Thus, an ideal interlayer chemistry is one that has high-mobility solutes that can also increase the strength in the joint region. A faster diffuser such as Zn may be viable, but a more compelling strategy is to explore solutes that promote precipitate formation. Both of these considerations can be explored computationally via CALPHAD approaches, using equilibrium calculations and mobility databases, and are ripe areas for future research.

# 4. Summary and Conclusions

This study evaluated the feasibility of TLP bonding to fabricate high-quality joints between superelastic NiTi alloys. Two TLP interlayer ternary systems, based on Cu and Nb solute additions, enabled effective wetting, dense joints, and complete isothermal solidification within 36 h at 1120 °C and 16 h at 1200 °C, respectively. The main conclusions are as follows:

(1) Isothermal solidification in both the Cu-base and Nb-base systems is limited by bulk diffusion, with the isothermal solidification distance exhibiting a $t^{1/2}$ dependence. At similar temperatures and equal times, the solidification distance of the Cu-base system is ~25% larger than that of the Nb-base system. When bonded at their respective minimum temperatures, the solidification distances are comparable over similar durations.

(2) TLP bonding produced a chemically uniform joint region, with maximum solute concentrations of 1.9 at.% Cu and 4.3 at.% Nb. The joint regions for both Cu-base and Nb-base bonds were found to be >98% B2 NiTi. $Ti_2Ni/Ti_4Ni_2O$ impurity phases were observed in small fractions, with 0.44% for the Cu-base bond and 1.18% for the Nb-base bond. This phase either formed due to unsolidified liquid from TLP, second-phase precipitation during cooling, or oxygen pickup during bonding.

(3) The TLP bonded specimens exhibited high joint efficiencies in terms of onset stress $\sigma_M^s$: 93.6% for the Cu-base bond and 88.8% for the Nb-base bond. Joint efficiencies based on UTS were 76.2% and 59.6%, respectively. Both bonds demonstrated stable superelastic behavior during cyclic loading, achieving recoverable strains of 3.70% (Cu-base) and 3.94% (Nb-base) in the 10th cycle when loaded to 425 MPa. After 10 cycles and subsequent loading to failure, the failure stresses of the Cu-base and Nb-base bonds were comparable to those obtained in the quasistatic tensile tests.

(4) DIC analysis revealed systematically lower strain amplitudes in the joint regions during cyclic loading, attributed to the higher elastic modulus from solute additions to NiTi, which was confirmed via nanoindentation.

# 5. Acknowledgements

This work is supported by NASA grant number ECF 80NSSC21K1810, and the Department of Defense through the NDSEG fellowship. GMV and AC acknowledge support from the National Science Foundation under award number ERC-2330175 for the Engineering Research Center EARTH. The authors thank Othmane Benafan, Santo Padula, and Travis Turner for the thoughtful discussions and feedback.

This work made use of the EPIC facility of Northwestern University's NUANCE Center, which has received support from the SHyNE Resource (NSF ECCS-2025633), the IIN, and Northwestern's MRSEC program (NSF DMR-2308691); and the MatCI Facility supported by the MRSEC program of the National Science Foundation (DMR-2308691) at the Materials Research Center of Northwestern University.

## Supplementary Materials for "Transient Liquid Phase Bonding of NiTi Using Cu- and Nb-base Interlayers"

Zhaoxi Cao[1], Samuel Price[1], Alessandra Crippa[2], John P. Reidy[1], Gianna M. Valentino[2], Ian McCue[1*]

[1]*Department of Materials Science and Engineering, Northwestern University, Evanston, IL 60208, USA*

[2]*Department of Materials Science and Engineering, University of Maryland, College Park, MD, 20742, USA*

*Corresponding author: ian.mccue@northwestern.edu*

***Table S1*** *TLP-relevant metrics determined via CALPHAD*

| Ternary Element | Temp (K) | $C_{x,\max}^{L\alpha}$ (at. %) | $\Omega_x^{\min}$ | $\Omega_x^{\max}$ | $K_x^{\min}$ | $K_x^{\max}$ | $D_x^{\min}$ | $D_x^{\max}$ |
|---|---|---|---|---|---|---|---|---|
| Ta | 900 | 4.13 | $7.41\text{x}10^{-2}$ | $2.79\text{x}10^{-1}$ | $4.00\text{x}10^{-2}$ | $1.35\text{x}10^{-1}$ | $1.53\text{x}10^{-16}$ | $2.97\text{x}10^{-16}$ |
| Mn | 1100 | 7.63 | $3.06\text{x}10^{-1}$ | $6.11\text{x}10^{-1}$ | $1.46\text{x}10^{-1}$ | $2.53\text{x}10^{-1}$ | $1.43\text{x}10^{-14}$ | $1.68\text{x}10^{-14}$ |
| Hf | 1200 | 6.59 | $1.41\text{x}10^{-1}$ | $6.68\text{x}10^{-1}$ | $7.29\text{x}10^{-2}$ | $2.70\text{x}10^{-1}$ | $4.57\text{x}10^{-14}$ | $1.29\text{x}10^{-12}$ |
| Nb | 1200 | 7.78 | 3.59 | 4.53 | $7.17\text{x}10^{-1}$ | $7.90\text{x}10^{-1}$ | $2.31\text{x}10^{-14}$ | $3.71\text{x}10^{-14}$ |
| V | 1200 | 1.02 | -1.58 | -1.58 | | | $7.63\text{ x}10^{-13}$ | $7.63\text{ x}10^{-13}$ |
| Zn | 1200 | 40.3 | $7.93\text{x}10^{-1}$ | 1.18 | $3.05\text{x}10^{-1}$ | $3.98\text{x}10^{-1}$ | $2.48\text{ x}10^{-13}$ | $2.77\text{ x}10^{-13}$ |
| Al | 1300 | 1.36 | -1.60 | -1.45 | | | $6.99\text{x}10^{-8}$ | $1.03\text{ x}10^{-7}$ |
| Co | 1300 | 19.7 | -1.42 | -1.05 | | | $1.05\text{x}10^{-13}$ | $8.20\text{x}10^{-13}$ |
| Cu | 1300 | 25.0 | $-3.77\text{x}10^{1}$ | $2.41\text{x}10^{1}$ | $1.76\text{x}10^{-1}$ | 1.30 | $1.59\text{x}10^{-13}$ | $7.90\text{x}10^{-13}$ |
| Fe | 1300 | 6.28 | -2.34 | -1.89 | | | $4.41\text{x}10^{-13}$ | $5.82\text{x}10^{-13}$ |
| Ir | 1300 | 2.21 | -1.03 | -1.02 | | | $1.59\text{x}10^{-13}$ | $1.92\text{x}10^{-13}$ |
| Mo | 1300 | 4.00 | $-4.17\text{x}10^{1}$ | -5.08 | | | $2.15\text{ x}10^{-14}$ | $6.76\text{ x}10^{-14}$ |
| Re | 1300 | 4.15 | -1.61 | -1.58 | | | $1.69\text{x}10^{-13}$ | $1.73\text{x}10^{-13}$ |
| Si | 1300 | 2.54 | -1.05 | -1.01 | | | $2.35\text{x}10^{-14}$ | $8.00\text{x}10^{-14}$ |
| W | 1300 | 1.75 | $1.72\text{x}10^{-3}$ | $2.29\text{x}10^{-2}$ | $9.70\text{x}10^{-4}$ | $1.28\text{x}10^{-2}$ | $2.65\text{x}10^{-14}$ | $3.10\text{x}10^{-14}$ |
| Zr | 1300 | 36.8 | -1.07 | 1.94 | $5.22\text{x}10^{-1}$ | $5.33\text{x}10^{-1}$ | $2.14\text{x}10^{-13}$ | $4.57\text{x}10^{-12}$ |

***Table S2.*** *Region-resolved DIC strain metrics during 25–425 MPa cyclic loading. Strains are averages over 700 µm windows positioned on the joint and on the upper and lower substrate regions.*

| Metric | Region | Cu-base bond | Nb-base bond |
|---|---|---|---|
| Peak strain, cycle 1 (%) | Joint | 1.12 | 1.45 |
| | Upper substrate | 4.65 | 4.64 |
| | Lower substrate | 5.15 | 4.03 |
| Residual strain after 10 cycles (%) | Joint | 0.06 | 0.18 |
| | Upper substrate | 0.44 | 0.44 |
| | Lower substrate | 0.62 | 0.39 |
| Substrate peak-strain asymmetry, cycle 1 → 10 (%) | - | 0.49 → 0.52 | 0.55 → 0.67 |
| Gauge-average residual strain after 10 cycles (%) | - | 0.35 | 0.30 |